# Cantera-Based Python Computer Program for Solving Steam Power Cycles with Superheating

Osama A. Marzouk

*[1]College of Engineering, University of Buraimi, Al Buraimi, Postal Code 512, Sultanate of Oman*

*Abstract*— One of the main sources of electricity generation is power plants that use water (steam) to rotate turbines, which drive large electric generators. The steam can be generated from renewable or non-renewable energy sources, such as geothermal energy and nuclear fuels. Having an analysis tool for modeling the performance of such steam power plants can greatly help in reaching optimum designs, leading to less fuel consumption, reduced pollution, and cheaper electricity. It is further advantageous if such modeling tool is free to access, does not require many inputs from the user, and gives results in a very short time. These remarks establish a motivation for the current study. This article documents a computer code written in the Python programming language for numerically analysing the main processes in a steam power cycle with superheating. The code utilizes built-in thermodynamic properties for water in the open-source software package "Cantera". A validation case with a benchmarking example in the literature using an independent source of water properties suggests that the developed code is correct. The code can be viewed as an extension to the Python examples for thermodynamic and power generation applications. Cantera can handle both subcritical and supercritical types of superheating. In the subcritical superheating, the steam absolute pressure does not exceed 220.9 bar. In the supercritical superheating, water becomes in a special condition called supercritical fluid, with absolute pressures above 220.9 bar.

*Keywords*—Cantera, Python, Code, Program, Superheating.

## I. INTRODUCTION

Modern computing resources, including hardware computers and software programs, play an important role in solving complex engineering or technology problems that can take much longer times and efforts if handled without the appropriate tools. Computational modeling enables rapid exploration of different designs, variables optimization, and performance testing in a safe virtual environment [1–6].

The specific field of application considered here is electric power generation using steam (hot water vapor) and steam turbine power plants, either those operating using fossil fuel [7], nuclear fuel [8], or concentrated solar power or solar thermal systems [9]. Combined-cycle gas turbine (CCGT) power plants use a combination of gas turbines (with the working fluid being hot combustion product gases instead of steam), as the higher-temperature section, and steam turbines, as the lower-temperature section [10]. Therefore, such power plants are within the scope of this article because they still used steam power in a part of their operation.

A preliminary design of a steam power plant involves the analysis of a steam vapor power thermodynamic cycle, which is a repeating sequence of processes applied to a circulating water flow as a heat transfer medium such that heat at one location (a boiler, a superheater, a reheater, or a heat recovery steam generator - HRSG) is converted into mechanical rotation in another location (the steam turbines) [11]. This energetic rotational power can then be used to drive large electric generators [12]; such that electricity is generated at a large scale for commercial use.

There are several versions of the steam power cycle. The basic steam cycle is idealized as having four consecutive processes, four devices (one device per process), and four water states (one state exists in the transition between two consecutive processes). The basic steam cycle can also be referred to as a saturated cycle or a saturated-steam cycle [13] because a main characteristic of this thermodynamic cycle is that water is heated in a boiler barely to convert the water from the liquid state into the gaseous state (saturated steam) by boiling. On the other hand, a superheated steam cycle (or a steam cycle with superheating) includes one modification where the formed saturated steam is further heated beyond the temperature threshold needed for its formation. The steam with such an extra heating stage is called superheated steam. A superheated steam cycle has two advantages over the saturated steam cycle [14].

The first advantage is an increase in the energy conversion efficiency (from heat to mechanical rotational energy). The second advantage is avoiding a situation where a lot of undesired liquid water droplets are formed inside the steam turbine as the saturated steam expands within the turbine and becomes colder. Such liquid water droplets can cause damage to the internal turbine elements [15] if they become excessive. Like the saturated steam cycle, the superheated steam cycle has 4 processes, 4 devices, and 4 water states. These are:

- *Process 1*: Compression in a pump, from state 1 to state 2
- *Process 2*: Heat addition in a boiler with a superheater, from state 2 to state 3
- *Process 3*: expansion in a turbine, from state 3 to state 4
- *Process 4*: condensation in a condenser, from state 4 back to state 1 (the initial state)

The current study considers the superheated-steam cycle and presents a developed computer code for its analysis under arbitrary user-defined conditions. Using the thermochemical library (package) Cantera, the user does not need to compute or insert physical properties of water/steam during the cycle, because these are computed internally using the freely-distributed Cantera library. The code is written in the Python programming language, which is a popular language that has been used in data science and machine learning, with many free libraries released to extend its capabilities [16].

The current study addresses a research gap, where the built-in examples that come with Cantera for thermodynamic applications do not include the problem of superheated steam cycle. The current study proposes a solution by providing a developed code here that is an extension of a Cantera built-in example code for analyzing non-superheated steam cycles. In the next section, an overview about Cantera is provided. Then, the proposed solver code is described. This is followed by applying the proposed solver to a benchmarking superheated steam cycle whose solution is available in the literature, for the purpose of testing the accuracy of the proposed solver. Then, a favorable feature in the proposed code is discussed, which is its ability to model both subcritical and supercritical steam cycles (because it is based on the software package Cantera, which can compute water properties in both subcritical and supercritical regimes). Also, some changes that were made in the original Cantera built-in computer code that led to the proposed code are highlighted.

Concluding remarks are provided after this, and finally two appendices are presented. The first appendix lists the lines of the Python code of the developed solver, and the second appendix gives a sample screen output as displayed to the user through running the proposed code.

## II. ABOUT CANTERA

Cantera is a software development project and a free open-source library with multiple modelling tools for problems involving chemical reactions, thermodynamic processes, and mass transport [17]. It can be used in the environment of the programming language Python, or the proprietary numerical analysis software Matlab [18]. Cantera can also be used in applications written in C and Fortran 90. The Cantera project is sponsored by NumFOCUS, which is a non-profit charity in the United States. The Cantera library is developed by a team of volunteers.

Cantera may be viewed as a reputable suite of tools, given that it has been successfully incorporated in various research studies [19–24].

Cantera comes with a number of useful examples, which demonstrate the functions that it can perform. The Python examples for the thermodynamics field include a solver for the saturated steam cycle [25]. However, there is no solver for the superheated steam cycle (as of January 18, 2023) [26]. This work reports a developed code that is an extension of the saturated cycle solver of Cantera such that it can solve superheated steam cycles while allowing the user to enter up to five input parameters to uniquely determine the corresponding superheated cycle, and then the code provides a summary of the predicted performance of the investigated steam cycle.

The developed code is in the form of a single Python file (.py). In order to use it, the user needs first to install the Cantera library into the Python version they use [27]. The code requires a Cantera version 2.5.0 or later. The Cantera version used here is 2.6.0. As of January 18, 2023, it is the latest released version of Cantera, which was released in May 2022.

## III. CODE DESCRIPTION

The content of the code is provided in Appendix A. It should be noted that lines starting with the hash sign (#), text appearing to the right of a hash sign, or text enclosed between triple quotes (""") in the code are comments only. They provide short explanations of the corresponding portion of the code they appear in.

These comments can be safely deleted without impacting the operation of the code. Some blank lines (totally blank) are present within the code. These can also be removed, although they help in organizing the code by providing aesthetic separators between various code segments.

The indentation of lines is important because of the way Python interprets line indentation and uses it to identify the level or section of each code line.

Table 1 lists the names of some code variables with their meaning. Most of the names and letter symbols in the code reflect common naming conventions in classical thermodynamics textbooks [28].

**TABLE I**
**SOME VARIABLES IN THE COMPUTER CODE AND THEIR MEANING**

| variable | meaning | unit |
|----------|---------|------|
| eta_cycle | Cycle overall efficiency | None |
| p | Absolute (not relative to the atmospheric pressure) pressure of water | $N/m^2$ |
| h | Absolute (not relative to the reference triple point state) specific enthalpy of water | J/kg |
| s | Absolute (not relative to the reference triple point state) specific entropy of water | J/kg.K |
| T | Absolute (not relative to the freezing temperature of fresh water at a standard atmospheric pressure) temperature of water | K |
| Q | Quality (also called dryness fraction) of water (when composed of a mixture of gas and liquid phases). A value of (1.0) means fully vapor, a value of (0.0) means fully liquid, intermediate values indicate a vapor-liquid mixture. | None |

It should be noted that the pressures are true pressures (absolute pressures), which means they are expressed relative to the minimum possible pressure, rather than being relative to the atmospheric (or ambient) pressure. This means that the pressure values here cannot be negative.

The code starts with two introductory comment lines between a pair of triple quotes, briefly stating the purpose of the code, and the minimum version of Cantera library needed.

Then the input parameters are decided. This part is supposed to be adjusted by the user based on desired operation conditions. These parameters are listed in Table 2.

**TABLE II**
**THE INPUT PARAMETERS DECIDED BY THE USER IN THE CODE**

| parameter | meaning | unit |
|-----------|---------|------|
| eta_pump | Pump efficiency (also called pump isentropic efficiency). As a special case, if eta_pump = 1.0, the compression process is described as 'isentropic', which means it is ideal or reversible (best case, with neglected losses). | None |
| eta_turbine | Turbine efficiency (also called turbine isentropic efficiency). As a special case, if eta_turbine = 1.0, the expansion process is described as 'isentropic', which means it is ideal or reversible (best case, with neglected losses). | None |
| p_max | Maximum absolute pressure of water during the entire cycle, which is the boiler pressure (between state 2 and state 3). Thus, p2 = p3 = p_max. | $N/m^2$ |
| p_min | Minimum absolute pressure of water during the entire cycle, which is the condenser pressure (between state 4 and state 1). Thus, p4 = p1 = p_min. | $N/m^2$ |
| T_max | Maximum absolute temperature of water during the entire cycle, which is at the boiler exit and turbine inlet (state 3). Thus, T3 = T_max. | K |

The definition of a function called (compress) comes next in the code. It virtually performs the compression process (from state 1 to state 2) and computes the amount of work energy per kg of water needed to complete this compression. In real operations, such work energy should be supplied from an external mover, such as an electric motor. This compression is treated as adiabatic, which means it happens without heat transfer to or from the water inside the pump.

Then, another function definition (function: expand) appears. This function virtually performs the expansion process in the turbine (from state 3 to state 4) and computes the amount of work energy per kg of water extracted from steam during the expansion. In real operations, such work energy is in the form of a rotating shaft power, which can be connected to an electric generator for producing electricity. The expansion is adiabatic; thus, it is assumed to happen without heat transfer to or from the water inside the turbine.

A third function definition (printState) exists after this. This function displays a formatted summary about each water state (states 1, 2, 3, and 4). However, it does not have a role in the analysis itself. The displayed summary for each state includes the temperature (in °C), the absolute pressure (in bar), the relative specific enthalpy (in kJ/kg), and the relative specific entropy (in kJ/kg.K). The last two quantities are relative to a reference state, such that the relative specific enthalpy and the relative specific entropy are exactly zero at this reference state, which is the pure liquid water at the triple point. The triple point corresponds to a specific temperature and pressure combination at which pure water can exist as a vapor, a liquid, and a solid together in equilibrium [29]. The triple point temperature is 0.01 °C, and the triple point pressure is 610.16 N/m$^2$. The reference specific enthalpy in Cantera for water is -15,970.81 kJ/kg, and the reference specific entropy in Cantera for water is 3.51998 kJ/kg.K. These reference values are not necessary in the thermodynamic analysis. Their role is to adjust the absolute specific enthalpy and the absolute specific entropy such that the displayed values, using the (printState) function, at each water state correspond to relative quantities having a zero value for a pure liquid water at the triple point, making them easier to comprehend. The state summary includes the water temperature expressed in the more-convenient unit (°C), rather than the specialized scientific unit of (K). The state summary also includes the absolute pressure, expressed in the more-convenient unit (bar), rather than the specialized scientific unit of (N/m$^2$). One bar is equivalent to 100,000 N/m$^2$. Since the (N/m$^2$) is relatively very small, using the larger unit (bar) makes the values easier to read. The state summary also includes the specific enthalpy relative to the reference triple point state, expressed in kJ/kg (rather than J/kg) and the specific entropy relative to the reference triple point state, expressed in kJ/kg.K (rather than J/kg.K). These larger units used in the summary are more appropriate to the range of the reported quantities.

Also, full names, such as "Temperature" and "Absolute Pressure", (rather than short symbolic letters like "T" or "p") are used in the state summary report. This avoids potential ambiguity in the state summary by having different interpretation for the same symbol.

The main part of the code that performs the actual modelling of the steam cycle appears in the code after the three previously discussed auxiliary functions. The steps of the steam cycle modelling are as follows:

1. A water species object (w) is established.
2. Reference specific enthalpy (h$_{Ref}$) and reference specific entropy (s$_{Ref}$) are defined as being at the triple point of liquid water.
3. The water conditions are set according to state 1 through enforcing the user-defined pressure (p1) and a zero-quality condition (no vapor content). The absolute specific enthalpy is obtained from the water object and stored as the variable (h1). It is needed at a later stage in the code to compute the rejected (wasted) heat per kg of water. A summary of state 1 is displayed on screen using the (printState) function.
4. The water conditions are set according to state 2 through simulating an adiabatic compression process in a water pump, until the user-defined pressure (p_max) is reached, subject to the user-defined pump efficiency (eta_pump). The absolute specific enthalpy at the new state 2 (after compression) is obtained from the water object and stored as the variable (h2). It is needed at a later stage in the code to compute the added heat per kg of water during the heat addition process (which includes both boiling and superheating). The amount of work energy needed per kg of water during this compression process is computed and stored as a variable (pump_work). A summary of state 2 is displayed on screen using the (printState) function.
5. The water conditions are set according to state 3 through simulating a constant-pressure heat addition process in a water boiler with a superheater, such that the user-defined temperature (T_max) is reached. This process including the superheating stage (not just converting liquid water into saturated steam). The absolute specific enthalpy at the new state 3 (after superheating) is extracted from the water object and stored as the variable (h3). It is used (with the previously stored value h2) to calculate the amount of heat energy needed per kg of water during this heat addition process, which is then stored as a variable (heat_added). A summary of state 3 is displayed on screen using the (printState) function.

6. The water conditions are set according to state 4 through simulating an adiabatic expansion process in a steam turbine, until the user-defined pressure (p_min) is reached, subject to the user-defined turbine efficiency (eta_turbine). The absolute specific enthalpy at the new state 4 (after expansion) is not obtained from the water object and stored as happened for h1, h2, and h3. This action is not necessary because no further changes are to be performed in the water object. The amount of work energy available per kg of water during this expansion process is computed and stored as a variable (turbine_work). There is no need to simulate the condensation process in a condenser, where any water vapor in the exhaust of the steam turbine is converted into a liquid phase such that the water returns back to A water species object (w) is established.

7. Reference specific enthalpy ($h_{Ref}$) and reference specific entropy ($s_{Ref}$) are defined as being at the triple point of liquid water.

8. The water conditions are set according to state 1 through enforcing the user-defined pressure (p1) and a zero-quality condition (no vapor content). The absolute specific enthalpy is obtained from the water object and stored as the variable (h1). It is needed at a later stage in the code to compute the rejected (wasted) heat per kg of water. A summary of state 1 is displayed on screen using the (printState) function.

9. The water conditions are set according to state 2 through simulating an adiabatic compression process in a water pump, until the user-defined pressure (p_max) is reached, subject to the user-defined pump efficiency (eta_pump). The absolute specific enthalpy at the new state 2 (after compression) is obtained from the water object and stored as the variable (h2). It is needed at a later stage in the code to compute the added heat per kg of water during the heat addition process (which includes both boiling and superheating). The amount of work energy needed per kg of water during this compression process is computed and stored as a variable (pump_work). A summary of state 2 is displayed on screen using the (printState) function.

10. The water conditions are set according to state 3 through simulating a constant-pressure heat addition process in a water boiler with a superheater, such that the user-defined temperature (T_max) is reached.

This process including the superheating stage (not just converting liquid water into saturated steam). The absolute specific enthalpy at the new state 3 (after superheating)

11. water object and stored as the variable (h3). It is used (with the previously stored value h2) to calculate the amount of heat energy needed per kg of water during this heat addition process, which is then stored as a variable (heat_added). A summary of state 3 is displayed on screen using the (printState) function.

12. The water conditions are set according to state 4 through simulating an adiabatic expansion process in a steam turbine, until the user-defined pressure (p_min) is reached, subject to the user-defined turbine efficiency (eta_turbine). The absolute specific enthalpy at the new state 4 (after expansion) is not obtained from the water object and stored as happened for h1, h2, and h3. This action is not necessary because no further changes are to be performed in the water object. The amount of work energy available per kg of water during this expansion process is computed and stored as a variable (turbine_work). There is no need to simulate the condensation process in a condenser, where any water vapor in the exhaust of the steam turbine is converted into a liquid phase such that the water returns back to

13. The initial state 1 (from state 4). A summary of state 4 is displayed on screen using the (printState) function.

14. The overall cycle efficiency is computed, by subtracting the pump work per kg of water from the turbine work per kg of water. This gives the net output work per kg of water. Dividing this net output work per kg of water by the needed heat for the boiler-superheater per kg of water gives the cycle efficiency. All three energy quantities (pump work, turbine work, boiler-superheater heat) are normalized such that they correspond to 1 kg of water.

15. An overall summary about the cycle is displayed to the user through a group of Python (print) functions. This general summary includes the estimated needed pump work energy per kg of water, the estimated produced turbine work energy per kg of water, the estimated net useful work energy per kg of water, the estimated needed heat per kg of water, and the estimated rejected (lost, unexploited) heat per kg of water during the condensation process. Finally, the cycle efficiency is displayed as a percentage.

## IV. RESULTS

### A. Benchmarking Case

To validate the solver code, it was applied for analysing a benchmarking steam power cycle with superheating, which was published in the open literature [30, 31]. The cycle in that reference study was analysed independently of the Cantera package for determining the properties of water. Instead, an online water properties calculator was used [32], which was made freely accessible by Spirax Sarco Limited (a large UK-based company that works in the field of steam systems). The mathematical model of the steam cycle is similar in the external reference and the solver code of this study, although the solver here is more flexible and allows for non-ideal compression in the pump and non-ideal expansion in the turbine. The external reference was limited to ideal (isentropic) compression and expansion.

The particular parameters for this benchmarking steam cycle are provided in Table 3.

**TABLE III**
**PARAMETERS FOR THE BENCHMARKING STEAM CYCLE**

| parameter | value |
|---|---|
| eta_pump | 1.0 |
| eta_turbine | 1.0 |
| p_max | $5 \times 10^6$ N/m$^2$ (50 bar) |
| p_min | 12,500 N/m$^2$ (0.125 bar) |
| T_max | 873.15 K (600 °C) |

The results from the solver code are close to those obtained in the reference study. In particular, the overall cycle efficiency computed by the presented solver and the reference study become identical if a common number of four significant digits is kept. This matching testifies to the validity of the solver code presented here. The comparison of some obtained cycle data and water properties are presented in Table 4.

**TABLE IV**
**TEST CASE RESULTS AND COMPARISON WITH AN EXTERNAL STUDY**

| quantity | value (here) | value (benchmarking) | unit |
|---|---|---|---|
| temperature before the pump | 50.29 | 50.2 | °C |
| temperature after the pump | 50.47 | 50.4 | °C |
| steam quality at turbine exit | 0.889556 | 0.890 | None |
| pump work per kg | 5.04 | 5.029 | kJ/kg |
| heat added per kg | 3,450.93 | 3,451.466 | kJ/kg |
| turbine work per kg | 1,336.99 | 1,337.140 | kJ/kg |
| heat rejected per kg | 2,118.98 | 2,119.355 | kJ/kg |
| cycle efficiency | 38.60 | 38.5955 | % |

### B. Sample Code Output

A sample screen output is given in Appendix B. It corresponds to the benchmarking case.

## V. DISCUSSION

There are two types of steam power cycles, depending on the maximum pressure and temperature that the superheated steam reaches during the cycle. One type is called subcritical steam cycle, where either the pressure or the temperature of the superheated steam is below a threshold called the critical value or critical point value. At the critical point, the liquid phase and the gaseous phase of water (or other pure substance in general) merge and become indistinguishable [33]. The critical temperature of water is 374.1 °C, and the critical absolute pressure of pure water is 220.9 bar [34], which are compatible with those provided by Cantera. The software package Cantera utilized here has the ability of handling both subcritical and supercritical steam cycles. This is a valuable feature in that modelling software, and thus in the computer solver presented here. The benchmarking case discussed here belongs to the subcritical type of steam power cycles, because the maximum absolute pressure is 50 bar, which is below the critical absolute pressure for water. Supercritical steam cycles tend to provide higher energy conversion efficiency (thus, better performance) than subcritical steam cycles [35]. However, supercritical power systems require more sophistication in the materials [36].

Compared to the original Cantera example code (rankine.py) for analysing saturated steam cycles, the following are some modifications made in the code described here for the superheated steam cycles, which extends the capability of the original one.

- The user can select a target temperature for superheated steam (instead of being limited to the saturated-steam limit).
- The user can select a target pressure for the turbine exit and condensation (instead of being limited to the saturated-steam limit).
- The state summary (using the printState function) was revised to display data in a more user-friendly and concise format. For example, the steam quality is displayed only for water states with liquid-vapor mixture, where this quantity becomes meaningful.
- A more-detailed cycle performance summary is displayed at the end, rather than only reporting the cycle efficiency.

These modifications, the documentation of the developed code, and its validation constitute the main contribution of this study.

## VI. Conclusion

A computer code written in the Python programming language was presented and explained. It is an extension of one of the examples for the software package Cantera for modelling thermodynamic processes and chemical reactions. The computer solver can analyse arbitrary steam power cycles with superheating, with either subcritical or supercritical type. The solver is in the form of a single Python (.py) file, which can be executed in a Python interpreter that has the Cantera library (version 2.5.0 or later) already installed.

The computer code consists of 87 lines (75 lines if not counting 12 blank separation lines, and 61 lines if further not counting other 14 comment lines). Its screen output has 42 lines (including 8 blank separation lines). It is a good demonstration of the power of computer-based analysis in engineering modelling, with a complete thermodynamic cycle analysis is performed and reported nearly instantly and precisely, saving a lot of time and effort.

The solver was validated through benchmarking with an example analysis case in the literature. The program code was provided and made available for use in this study as an appendix, with another appendix showing a sample output from it. The solver can be helpful for engineers, researchers, or students involved in any of the fields of fuel-fired steam power plants, combined cycle power plants, nuclear power plants, solar thermal power plants, thermodynamics, computational modelling, or Python programming.

*APPENDIX A: PROGRAM CODE LINES*

```
"""

Python solver for steam vapor power cycles with superheating.

It requires Cantera package version 2.5.0 or later.

"""

import cantera as ct

# parameters

eta_pump = 1.0    # pump isentropic efficiency (fraction, not percentage)

eta_turbine = 1.0 # turbine isentropic efficiency (fraction, not percentage)
```

```
p_max = 50.0e5          # maximum pressure, absolute, in Pa (N/m2)

p_min = 0.125e5 # minimum pressure, absolute, in Pa (N/m2)

T_max = sum([600 , 273.15])     # maximum temperature, in K (absolute)

def compress(fluid, p_final, eta):
    """Adiabatically compress a fluid to pressure p_final, using a pump with an isentropic efficiency eta"""
    h0 = fluid.h
    s0 = fluid.s
    fluid.SP = s0, p_final
    h1s = fluid.h
    isentropic_work = h1s - h0
    actual_work = isentropic_work / eta
    h1 = h0 + actual_work
    fluid.HP = h1, p_final
    return actual_work

def expand(fluid, p_final, eta):
    """Adiabatically expand a fluid to pressure p_final, using a turbine with an isentropic efficiency eta"""
    h0 = fluid.h
    s0 = fluid.s
    fluid.SP =s0, p_final
    h1s = fluid.h
    isentropic_work = h0 - h1s
    actual_work = isentropic_work * eta
    h1 = h0 - actual_work
    fluid.HP = h1, p_final
    return actual_work

def printState(n, fluid):
    print('\n*************** State {0} ****************'.format(n))
    print('Temperature (degree Celsius) = {:6.2f}'.format(fluid.T - 273.15))
    print('Absolute pressure (bar) = {:.3f}'.format(fluid.P/1e5))
    print('Specific enthalpy (kJ/kg) = {:.4f}'.format((fluid.h-hRef)/1000))
    print('Specific entropy (kJ/kg.K) = {:.6f}'.format((fluid.s-sRef)/1000))
    x = fluid.Q # quality (dryness fraction), from 0 to 1
    if x < 1 and x > 0:
        print('Liquid-gas mixture, quality = {:.6f}'.format(x))
```

```
# The main analysis starts here
w = ct.Water() # make an object representing water

# reference state (triple point; liquid state)
w.TQ = w.min_temp, 0
hRef = w.h
sRef = w.s

# Start with saturated liquid water at p_min
w.PQ = p_min, 0.0
h1 = w.h
p1 = w.P
printState(1, w)

# Process 1 (in pump): Compress water adiabatically to p_max
pump_work = compress(w, p_max, eta_pump)
h2 = w.h
printState(2, w)

# Process 2 (in a boiler): Add heat to water at a constant pressure until it reaches the desired superheated steam state
w.TP = T_max, p_max
h3 = w.h
heat_added = h3 - h2
printState(3, w)

# Process 3 (in a steam turbine): expand water back to the initial pressure p1
turbine_work = expand(w, p1, eta_turbine)
printState(4, w)

# efficiency
eta_cycle = (turbine_work - pump_work)/heat_added

print('\n***************** Cycle Analysis *****************')
print('Input pump work per kg (kJ/kg) = {:.2f}'.format(pump_work/1000))
print('Output turbine work per kg (kJ/kg) = {:.2f}'.format(turbine_work/1000))
print('\nNet output work per kg (kJ/kg) = {:.2f}'\
    .format((turbine_work-pump_work)/1000))
print('Input heat per kg (kJ/kg) = {:.2f}'.format(heat_added/1000))
print('\nHeat rejected per kg (kJ/kg) = {:.2f}'.format((w.h-h1)/1000))
print('\nCycle efficiency = {:.2f}%'.format(100* eta_cycle))
```

*APPENDIX B: SAMPLE OUTPUT OF THE PROGRAM CODE*

***************** State 1 ******************

Temperature (degree Celsius) =  50.29

Absolute Pressure (bar) = 0.125

Specific enthalpy (kJ/kg) = 210.5134

Specific entropy (kJ/kg.K) = 0.707436

***************** State 2 ******************

Temperature (degree Celsius) =  50.47

Absolute Pressure (bar) = 50.000

Specific enthalpy (kJ/kg) = 215.5568

Specific entropy (kJ/kg.K) = 0.707436

***************** State 3 ******************

Temperature (degree Celsius) = 600.00

Absolute Pressure (bar) = 50.000

Specific enthalpy (kJ/kg) = 3666.4839

Specific entropy (kJ/kg.K) = 7.258869

***************** State 4 ******************

Temperature (degree Celsius) =  50.29

Absolute pressure (bar) = 0.125

Specific enthalpy (kJ/kg) = 2329.4947

Specific entropy (kJ/kg.K) = 7.258869

Liquid-gas mixture, quality = 0.889556

***************** Cycle Analysis ******************

Input pump work per kg (kJ/kg) = 5.04

Output turbine work per kg (kJ/kg) = 1336.99

Net output work per kg (kJ/kg) = 1331.95

Input heat per kg (kJ/kg) = 3450.93

Heat rejected per kg (kJ/kg) = 2118.98

Cycle efficiency = 38.60%